# Electrical properties of PbS films doped with iodine by chemical bath deposition

T.B. Charikova[1], A.Yu. Pavlova[1], M.R. Popov[1], A.V. Pozdin[2], L.N. Maskaeva[2,3]

[1]*M.N. Mikheev Institute of Metal Physics of Ural Branch of Russian Academy of Sciences, 18 S. Kovalevskaya Str., Ekaterinburg, 620108, Russia;*

[2]*Ural Federal University Named After the First President of Russia B.N. Yeltsin, 19 Mira Str., Ekaterinburg, 620002, Russia;*

[3]*Ural Institute of State Fire Service of EMERCOM of Russia, 22 Mira str., Ekaterinburg, 620062, Russia.*

**ABSTRACT**

We present the results of measurements of bulk current-voltage (I-V) characteristics and local surface I-V characteristics by atomic force microscopy (AFM) of iodine-doped PbS films. It is established that bulk I-V curves of both undoped and iodine-doped PbS films demonstrate a linear (ohmic) *U*(*I*) dependence. The type of local surface I-V characteristics is ohmic at the concentration range of the dopant $0 < [NH_4I] \le 0.10$ M and becomes rectifying at $[NH_4I] \ge 0.15$ M, which is determined by a decrease in the size and a change in the shape of the film grains, as well as a decrease in the surface roughness of the film. An increase in the iodine content in the PbS(I) films leads to nonlinear dependences of the microscopic characteristics and photoelectric parameters of the PbS(I) films. A sharp decrease in the diffusion coefficient, the beginning of an increase in the charge carrier lifetime, a maximum in voltage sensitivity and specific detectability are observed in the PbS(I) film chemically deposited from a reaction mixture containing $[NH_4I] = 0.15$ M. This indicates that the optimal concentration of iodine in the film is 2.7 at.%.

## INTRODUCTION

Infrared photodetectors (IRPDs) currently have a very wide range of applications. They are used ~~for~~ in night vision devices, communication systems, medical imaging, atmospheric sensing, environmental monitoring, in spectral scientific instruments, in space research and astronomical observations [1-4]. The range of application of IR photodetectors is constantly expanding. Lead sulfide, which operates on the internal photoelectric effect, is a promising narrow-band $A^{IV}B^{VI}$ semiconductor for recording IR radiation as an active element. The development of infrared technology is closely related to the study of phenomena both in the volume and on the surface of semiconductors. Therefore, knowledge of the compounds included in the bulk and surface of PbS films formed by chemical deposition is very important for processes control. Despite the development of serial technologies for the IRPDs production, the problems to fabricate a thin (1-2 μm) homogeneous semiconductor film with a minimum free carriers' concentration (it is desirable to obtain min $p = (1-3) \cdot 10^{15}$ cm$^{-3}$ at $T = 300$ K) and to adapt photodetectors to the required for the optical-electronic systems frequency range remain unresolved.

Photoelectric phenomena and recombination processes in $A^{IV}B^{VI}$ compounds and, in particular, in PbS have been studied for several decades. Measurements of the Hall coefficient and electrical conductivity have been widely used to determine the mobility and concentration of carriers in PbS films under various temperature conditions [5]. By the middle of the last century, the existing theories could be divided into three directions: the concentration modulation theory (proposed by Moss [6], Bode [7] and Livenshtein [8]), the barrier theory (proposed by Gibson [9] and Slater [10]) and the generalized Petritz theory [11]. Later, Neustroev and Osipov [12, 13] proposed an explanation of the electrical and photovoltaic properties of polycrystalline $A^{IV}B^{VI}$ layers. According to this model, the layers consist of n-type crystallites, and on the surface of this crystallites there are *p*-type inversion layers associated with acceptor states in the intercrystalline layers.

An important feature of semiconductors is the significant dependence of their electrical properties on the surface state and doping with both cations [14] and anions. The real surface of metal sulfides is usually depleted of metal, while distortions of the sulfide structure up to amorphization can capture the volume of the solid to a significant depth of up to several micrometers [15]. Distortions of the structure in the surface layer, the appearance of adsorbed particles on the surface of the film, other defects of the surface and surface layer lead to the appearance of local energy levels in the energy band gap of the semiconductor, curvature of the bands near the film surface and a change in the mechanism and type of conductivity.

Analysis of the current-voltage characteristics of semiconductor structures provides an information on the features of charge carrier transport [16-18]. The I-V characteristic may vary depending on conditions such as temperature and illumination. If lead sulfide is used in contact with a metal, a Schottky barrier may form at the interface. This barrier affects the shape of the current-voltage characteristic, especially at low voltages. In this case, the current-voltage characteristic may exhibit rectifying properties. In real PbS-based devices, uneven distribution of charge carriers is possible, which may also affect the shape of the current-voltage characteristic. The aim of our work was to identify the relationship between the main parameters of photoreceiving devices based on PbS doped with iodine (voltage sensitivity $U_s$, time constant $\tau$ and specific detectability $D^*$) and the microscopic parameters of the solid – the concentration of the main charge carriers and the lifetime of free charge carriers, as well as features of the film surface layer as a result of the analysis of electrophysical and volt-ampere characteristics.

**EXPERIMENTAL METHODS**

1. **Sample preparation procedure**

To obtain PbS films, aqueous solutions containing 0.04 M lead acetate $Pb(CH_3COO)_2$, 0.30 M sodium citrate $Na_3C_6H_5O_7$, 4.0 M aqueous ammonia $NH_3 \cdot H_2O$ and 0.58 M thiourea $N_2H_4CS$ were used. Films obtained by additionally introducing 0.05 to 0.25 M ammonium iodide into the reaction solution are designated as PbS(I). The qualification of the reagents used is "chemically pure". Chemical deposition was carried out in sealed reactors on pre-defatted silicate glass substrates (72.2% $SiO_2$, 14.3% $Na_2O$; 1.2% $K_2O$, 6.4%, 4.3% MgO, 1.2% $Al_2O_3$, 0.03% FeO, 0.3% $SO_3$) at a temperature of 353 K in a TS-TB-10 thermostat with a temperature maintenance accuracy of ±0.1 °C. The duration of thin-film layer deposition was 90 minutes.

The thickness of the synthesized films was estimated using a Linnik MII-4M microinterferometer, providing a measurement accuracy of ±10%.

Surface morphology and the content of the main elements (Pb, S, I) in thin-film lead sulfide were determined by scanning electron microscopy (SEM) using a Tescan Vega 4 LMS microscope with energy-dispersive X-ray spectroscopy (EDS) Oxford Xplore EDS – AZtecOne. The particle size distribution on the film surface was assessed using Image J and Origin software.

2. **Experimental methods**

In our previous work [19], we studied the topography and surface roughness of PbS films doped with iodine in the chemical deposition process, conducted X-ray structural studies, obtained vibrational spectra of the synthesized films, and measured electrical resistivity and Hall resistivity.

In this work, standard methods were used to obtain the I-V characteristics of the film bulk and the atomic force microscopy (AFM) method was used to record local surface I-V curves of iodine-doped PbS films. The Solver Next (NT-MDT) atomic force microscope and NSG01 (NT-MDT) n-type silicon cantilevers with conductive 20-30 nm Pt coating, 35 nm tip curvature radius and force constant of 1.45-15.1 N/m were used in the contact mode to study local electrical characteristics of films such as spreading resistance and local I-V curves. The sample surface was grounded by the spring contact and constant bias voltage $U$ was applied to the tip. The resulting tip-sample current $I_{pr}$ was measured as a function of the probe scanning position and AFM surface spreading current images were obtained at different $U$ values varied from -3 V to +3 V with 0.5 V step. As tip-sample contact resistance is supposed to be constant, the current was proportional to the local sample resistance. Then the tip-sample current $I_{pr}$ was measured as a function of the applied tip-sample voltage $U$ varied from -10 V to +10 V with 0.02 V step and local I-V curves were obtained for 25-30 points on the film surface. All AFM measurements were carried out in a sample scanning configuration under ambient conditions in a clean room with a controlled temperature of 20 °C and relative humidity of 25-30%. The 5x5 µm² areas were imaged at scanning rate of 0.5-1 Hz with resolution scanning step of 10 nm. Loading force of the probe was the same for all studied samples and sufficient for stable contact between the probe and the surface, but not leading to their destruction. Subsequent analysis of AFM images was done using Nova Image Analysis (NT-MDT) and Gwyddion software.

The 5x5 mm² sized sensitive elements with electrochemically deposited ohmic nickel contacts were fabricated to investigate the photosensitive properties of the films. The measurements were done on a specialized stand K.54.410 with absolutely black body 573K as a radiation source at $1·10^{-4}$ W/cm² irradiance of the sensitive element, 800 Hz modulation frequency and 25 V bias voltage. The photoresponse of the elements was recorded as a voltage drop across a matched load resistance. The time constant (τ) was recorded using an ISDS205B digital oscilloscope when the output signal reached 0.7 of the maximum value. The specific detectability (D*) of the sensitive elements was determined by calculation based on measurements of the photoresponse and noise voltage. The following expression was used to calculate D*:

$$D^* = \frac{U_S \cdot \sqrt{\Delta f}}{U_n \cdot E_{\text{eff}} \cdot \sqrt{A_p \cdot K_u}} \qquad (1)$$

Where $U_S$ is the photoresponse magnitude, $\Delta f$ is the frequency bandwidth of amplification path ($\Delta f = 240$ Hz), $U_n$ is the noise voltage of receiver, $E_{\text{eff}}$ is the effective density of the radiation flux on the sensitive element ($E_{\text{eff}} = 0.25 \cdot 10^{-4}$ W/cm²), $A_p$ is the area of the sensitive photoreceiving element ($A_p = 0.25$ cm²), $K_u$ is the radiation utilization coefficient ($K_u = 0.1$).

## RESULTS AND DISCUSSION

The analysis of the temperature dependencies of the electrical resistivity and the Hall resistivity of polycrystalline lead sulfide films doped with iodine carried out in previous work [19], showed that the type and concentration of charge carriers change depending on the dopant content in the layer. The temperature dependences of the Hall constant and the mobility of charge carriers have a complex nature which is explained not only by the properties of the crystallites but also of the intercrystalline layers (the composition and properties depend on the presence of microimpurities and inclusions carried to the interphase boundary during the crystallization process), the multiphase composition of the films and the possibility of the occurrence of p-n junctions at the interfaces. Thus, the obtained activation dependence of the electrical resistance for the studied films in the temperature range of 350–170 K can be explained by the potential barrier between crystallites. When ammonium iodide is introduced into the reaction solution and iodine is included in the composition of the films the increase of the amount of point defects and dislocations, and misorientation of grains in the layer is occurred. This process is accompanied by a change of conductivity from electron type (n-type) in undoped PbS films to hole type (p-type) with an increase in the concentration of the halide dopant in the reaction mixture and, as a consequence, the concentration of the main carriers. The explanation of this results seems to be the most reasonable by the self-compensation mechanism for the donor action of halogens, previously established for metal chalcogenides [19-22].

The thickness of the fabricated PbS films on glass substrates decreases from 450 to 200 nm with an increase in the concentration of ammonium iodide in the reaction solution from 0 to 0.25 M. According to the electron micrographs of the thin-film lead sulfide, a significant influence of the iodide additive on the size, shape and orientation of the grains forming the semiconductor compound is observed [23]. Deposited films have different microstructure and grain size distribution is unimodal. The undoped PbS film have well-faceted and close-packed crystallites with 400-800 nm average size. The presence of minimal $NH_4I$ concentration (0.05 M) in reactor leads to decrease in the grain size to 200-500 nm without change in the grain shape. Also 3% of nanoparticles appear at this concentration, their number increases to 7% at 0.10 M of $NH_4I$ and reaches a maximum 10% with a loss of grain faceting at 0.15 M of $NH_4I$. A further increase in dopant concentration leads to decrease in the amount of nanoparticles to 2% (at 0.20 M) and their absence at the maximum ammonium iodide concentration 0.25 M. Thus, the $NH_4I$ concentration of 0.15 M can be called critical to the PbS(I) films formation.

As a result of introducing 0.05 to 0.25 M ammonium iodide into the reaction solution, the grain shape changes and the grain size decreases with the increase in the amount of nanoparticles and uniform layer microstructure forms [19, 23].

The content of the main elements (Pb, S, I) in thin-film lead sulfide was determined by EDS microanalysis of the whole film surface. On Fig. 1 there are SEM images of the analyzed areas and EDX spectra.

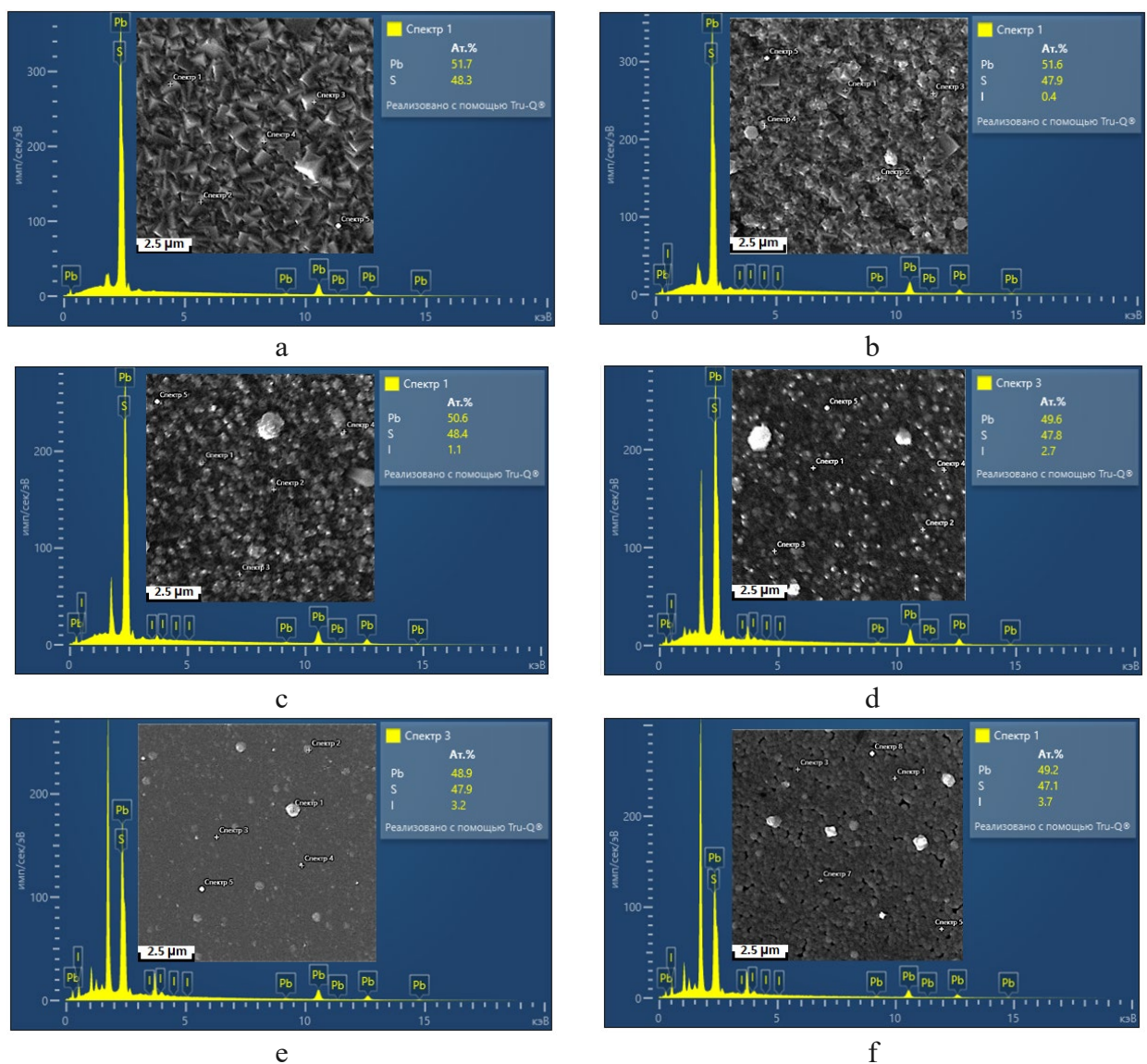

FIG. 1. SEM images of the film surface for PbS undoped (a) and iodine-doped PbS (b-f) indicating the areas where EDX analysis and spectra were performed. Concentration of $NH_4I$, M: 0.05 (b), 0.1 (c), 0.15 (d), 0.2 (e), 0.25 (f).

Undoped PbS film contained 51.4 at.% Pb and 48.6 at.% S and iodine doped PbS films contained 48.9-50.8 at.% Pb and 47.3-48.2 at.% S, i.e. there is a slight excess of metal over chalcogen. Moreover, the Pb/S ratio decreases from 1.11 to 1.03 with an increase in the iodine content from 0.4 to 3.2-3.7 at.%.

As it was noticed before, undoped PbS films have n-type conductivity. Deposition of PbS films from solutions containing $NH_4I$ changes their conductivity to p-type.

On Fig. 2 I-V characteristics of undoped and iodine-doped PbS films measured by standard 4-probe method at room temperature are shown. Pure PbS film and PbS film with 0.04 at.% I demonstrate a linear (ohmic) U(I) dependence in the current range from -10 mkA to +10 mkA. I-V characteristics of PbS films with 3.2 and 3.7 at.% I show deviation from linearity at currents I ≥ 2 mkA. This behavior indicates the presence of concentration sufficient to provide thermally activated conductivity at room temperature (Table 1).

As can be seen from Table 1, the dark specific resistance $\rho$ increases more than two orders of magnitude with an increase of iodine content to 3.7 at.% while introduction of iodine into the film and an increase of its content to 3.2 at.% change the conductivity to the hole-type. The charge carriers concentration decreases and reaches a minimum of $8.5\times10^{15}$ cm$^{-3}$ at iodine content 3.2 at.% and their mobility increases by 1.8 times and reaches 33.57 cm$^2$/V·s with an increase in the iodine content to 1.1 at.%. The mobility begins to decrease sharply at iodine content of 2.7 at.% and at 3.7 at.% decreases by two orders of magnitude reaching 0.03 cm$^2$/V·s.

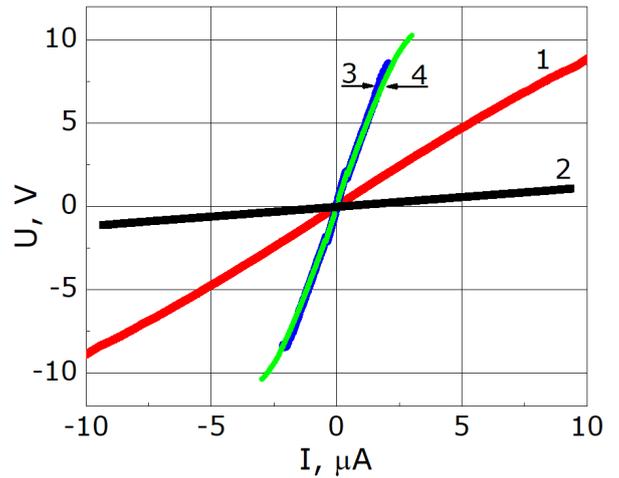

Fig. 2. I-V characteristics of PbS films (1) and PbS films with 0.4 at.% (2), 3.2 at.% (3), 3.7 at.% (4) of I, measured at T = 300 K.

Apparently, as was noticed in [19], the introduction of donor iodine dopant into the PbS film leads to the formation of acceptor vacancies of Pb and as a consequence to an increase in the electron concentration. However, when the iodine content is greater than 2.7 at.%, the increase of acceptor vacancies suppresses the electron concentration growth and leads to the equalization of the concentration of both types of charge carriers, an increase in the hole concentration and a change in the conductivity type. The transition to the hole conductivity leads to a decrease in the charge carriers mobility which, in turn, depends on the decrease in the film crystallites sizes with an increase in the iodine content.

TABLE 1. Dark specific resistance $\rho$, charge carriers concentration $n(p)$ and mobility $\mu$ for undoped PbS and iodine-doped PbS films at $T = 300$ K.

| Concentration of [NH$_4$I], M | I, at. % | $\rho$, $\Omega \cdot$m (T = 300 K) | $n(p) \cdot 10^{17}$, cm$^{-3}$ | $\mu$, cm$^2$/V$\cdot$s |
|---|---|---|---|---|
| 0 | 0 | 0.011 | -4.14 | 18.32 |
| 0.05 | 0.4 | 0.116 | -0.24 | 22.07 |
| 0.1 | 1.1 | 0.175 | -0.11 | 33.57 |
| 0.15 | 2.7 | 1.03 | -1.69 | 0.36 |
| 0.2 | 3.2 | - | +0.08 | |
| 0.25 | 3.7 | 3.134 | +6.9 | 0.03 |

Fig. 3 shows AFM topography and surface spreading current images at different bias voltages on the PbS(I) film, chemically deposited at minimal concentration of ammonium iodide (0.05M). There is some difference of image areas because of the scanning drift. When the small bias voltage is applied, the current starts to flow on the grain boundaries and at $U = 2$ V current is distributed homogenously across the grains of the sample.

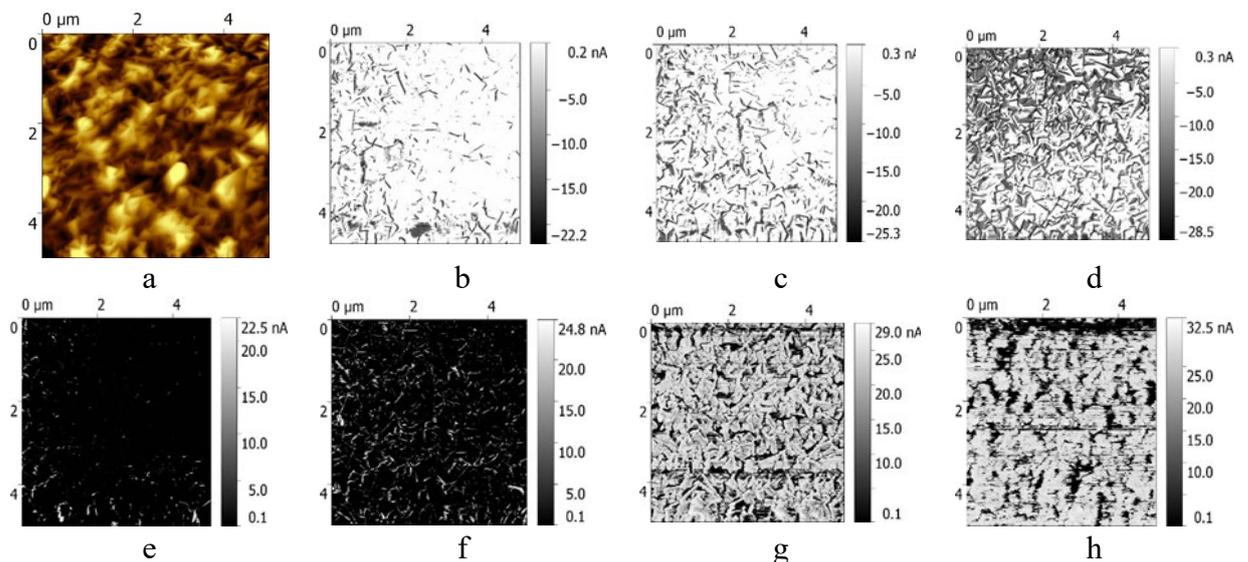

Fig. 3. AFM topography (a) and surface spreading current (b-h) images for the PbS(I) film, containing 0.4 at.% I. Current images are obtained at different bias voltages $U$: -1 V (b), -2 V (c), -3 V (d), 0.5 V (e), 1 V (f), 2 V (g) and 3 V (h).

Profiles, extracted from spreading current images, show that the largest increase in conductance occurs at positive bias voltage 2 V and negative bias voltage -3 V (Fig. 4).

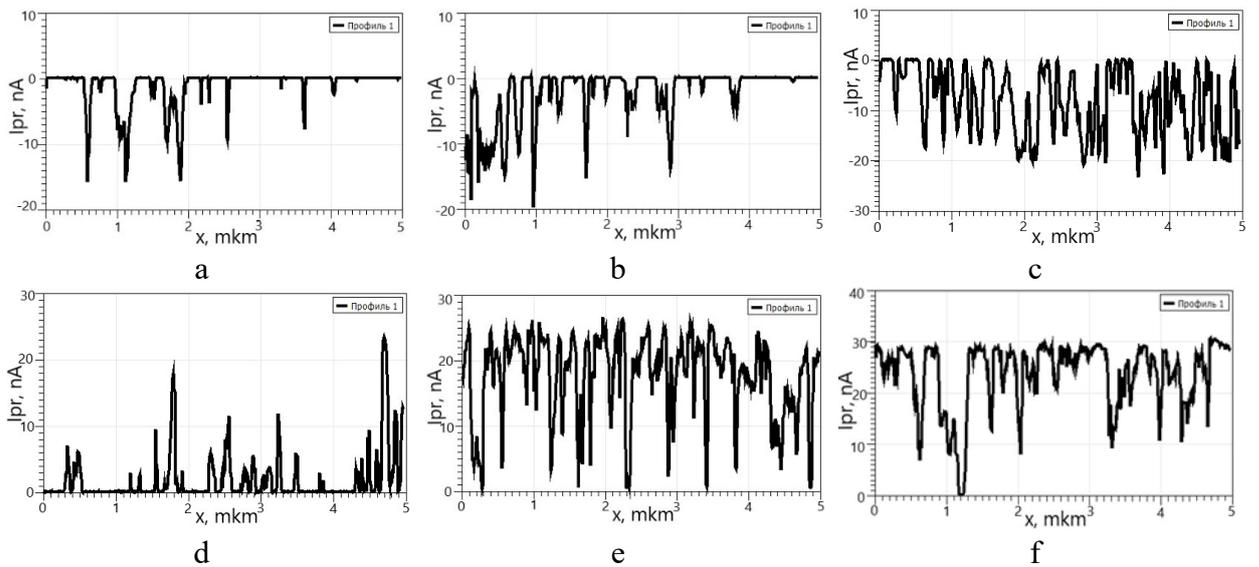

Fig. 4. AFM surface spreading current profiles for the PbS(I) film, containing 0.4 at.% I. Profiles are obtained at different bias voltages *U*: -1 V (a), -2 V (b), -3 V (c), 1 V (d), 2 V (e) and 3 V (f).

For pure PbS and PbS(I) films, containing from 0.4 at.% to 2.7 at.% I, currents measured at positive bias voltages are slightly higher than at the negative ones. But at higher iodine contents (3.2 and 3.7 at.%) currents for both positive and negative bias become almost the same. In general, the surface spreading current decreases for doped PbS(I) films with an increase in iodine content (Fig. 5), which indicates an increase in local surface resistance. This fact is confirmed by the results of measurements of local I-V characteristics.

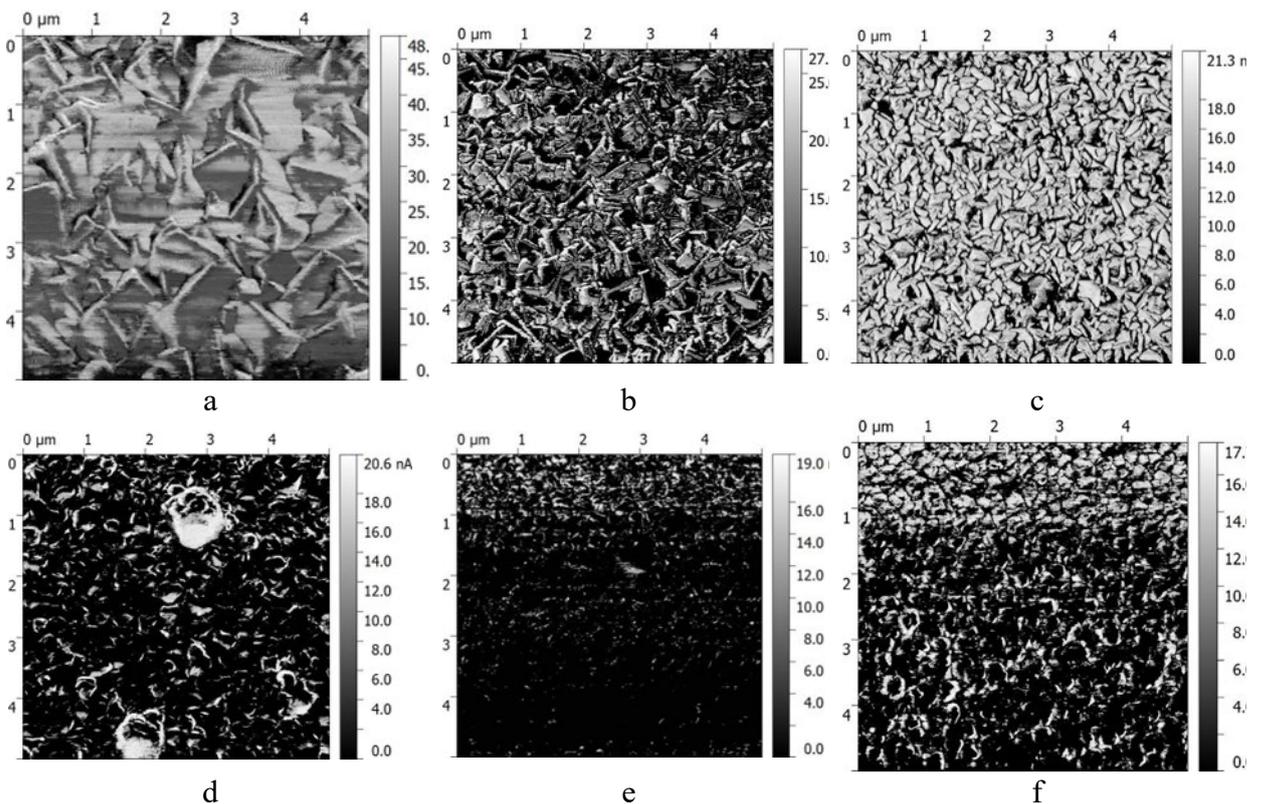

Fig. 5. AFM surface spreading current images for the pure PbS film (a) and PbS(I) films, containing I, at.%: 0.4 (b), 1.1 (c), 2.7 (d), 3.2 (e) and 3.7 (f) at bias voltage $U = +1.5$ V.

Fig. 6 shows single I-V characteristics from a series of local obtained by AFM dependences $I_{pr}(U)$ for undoped PbS film and iodine-doped PbS(I) films. It should be noted that these dependencies are typical for the I-V curves measured using a probe with a conductive 20-30 nm Pt coating under "clean room" conditions (see the "Experimental methods" section). In general, the local I–V characteristics obtained for both undoped PbS and iodine-doped PbS(I) films differed at various points on the surface, which is due to its inhomogeneity.

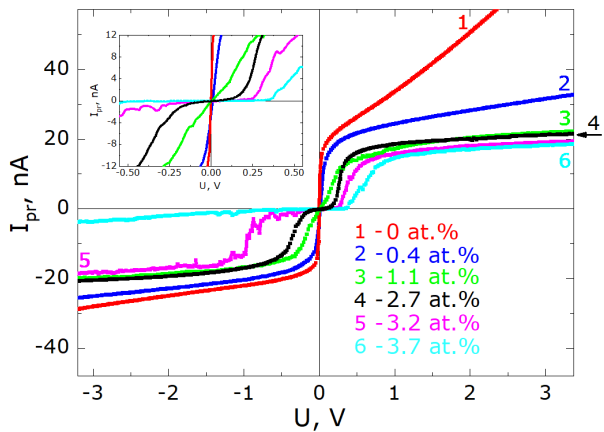

Fig. 6. Local I-V characteristics for the pure PbS film (1) and PbS(I) films, containing I, at.%: 0.4 (2), 1.1 (3), 2.7 (4), 3.2 (5) and 3.7 (6) at $T = 300$ K. The inset shows I-V characteristics in the bias range -0.5 V $< U <$ +0.5 V.

A change in the behavior of local I-V curves is observed when the level of iodine doping of PbS films changes. For an undoped PbS film, the local I-V characteristic has an ohmic appearance, and at $U = 30$ mV, a change in the slope of the ohmic I-V characteristic occurs, which indicates a change in the film resistance. This situation is also observed for PbS(I) films obtained from reaction mixtures containing 0.4 and 1.1 at.% I at a bias voltages $U$ equal to 90.6 and 271.9 mV respectively. However, the $I_{pr}(U)$ dependencies for PbS(I) films containing $\geq$ 2.7 at.% I have a clear asymmetric behavior. The current grows exponentially with the applied voltage in the forward bias range ($U < +0.5$ V). In the reverse bias range ($U > -0.5$ V), the current is almost independent of the voltage. Similar rectifying current-voltage characteristics on Cu-doped PbS films were presented in [14]. Problems of metal-semiconductor contact are discussed in [14, 24-27] due to the features of local I-V characteristics on films surface. In an idealized model [24] that neglects Fermi level pinning and other surface effects, the authors claim that such a contact should be rectifying if the semiconductor is n-type and ohmic if it is p-type. However, the results of the I-V measurements presented in [14, 25, 26] show that the contact is rectifying, even if the semiconductor has p-type conductivity. In our measurements, the rectifying properties of the contact appear at a level of iodine doping of the PbS film equal to 2.7 at.% (n-type), and with a further increase in the doping level, a change in the sign of the main charge carriers occurs.

The rectifying contact of metal with n-type semiconductor (Schottky contact) occurs when the work function of the metal ($A_M$) is greater than the work function of the n-type semiconductor ($A_{SN}$). The rectifying contact of metal with p-type semiconductor is formed when the work function of the metal is less than the work function of the electrons from the p-type semiconductor ($A_{SP}$). Taking into account these facts it can be stated that for the studied films $A_M < A_{SN}, A_{SP}$. The work function of a semiconductor depends not only on the width of the band gap, but also on the doping and surface state. Indeed, as we showed in [19], at a level of doping of the PbS film with 2.7 at.% I, the film surface becomes more uniform and less rough, the grain size decreases to 200-250 nm and the proportion of nanoparticles increases to a maximum (up to 10%). For PbS films doped with 2.7, 3.2 and 3.7 at.% I the threshold voltages of the rectifying contact $U_0$ are small and equal to 50.3 mV, 151.0 mV and 292.0 mV respectively. Rectifying contacts that appear on the surface of PbS(I) films containing 2.7, 3.2 and 3.7 at.% I are explained, as was mentioned above, by the properties of the intercrystalline layers and the possibility of the occurrence of p-n junctions at the interfaces [19].

An increase in the number of both point defects and dislocations, grain misorientation in the layer upon the introduction of ammonium iodide into the reaction solution and the inclusion of iodine in the composition of the films is accompanied by a change in the conductivity type and the charge carriers concentration (see Table 1). For the above-mentioned reasons, when the concentration of n(p) charge carriers in a semiconductor is non-uniform, diffusion currents occur.

The correlation between the diffusion coefficient and the mobility of charge carriers is expressed by Einstein's relation:

$$D = \frac{kT}{e}\mu = \varphi\mu \qquad (2),$$

where $k$ is the Boltzmann constant, $T$ is the temperature, $e$ is the electron charge, $\varphi = kT/e$ is the thermal potential, $\mu$ is the charge carriers' mobility.

The average diffusion length of charge carriers (diffusion length) is determined by their lifetime and diffusion coefficient:

$$L = \sqrt{\tau_e D} \qquad (3)$$

TABLE 2. Photoelectric parameters (voltage sensitivity $U_s$, time constant $\tau$, specific detectability D*) and microscopic characteristics of the solid (diffusion coefficient $D$ and charge carriers' lifetime $\tau_e$) depending on iodine content in PbS films.

| [I], ат. % | 0 | 0.4 | 1.1 | 2.7 | 3.2 | 3.7 |
|---|---|---|---|---|---|---|

| $U_s$, мкВ | 50 | 300 | 700 | 2300 | 1300 | 510 |
|---|---|---|---|---|---|---|
| $\tau$, мкс | 35 | 42 | 70 | 210 | 400 | 585 |
| $D^*$, см·Гц$^{0.5}$·Вт$^{-1}$ | $1.9\cdot10^8$ | $1.1\cdot10^9$ | $3.8\cdot10^9$ | $8.3\cdot10^9$ | $5.7\cdot10^9$ | $1.9\cdot10^9$ |
| $D$, см$^2$·с$^{-1}$ | 0.47 | 0.57 | 0.84 | $9.3\cdot10^{-3}$ | - | $7.5\cdot10^{-5}$ |
| $\tau_e$, мкс | 0.021 | 0.018 | 0.012 | 1.074 | - | 13.326 |

Where $\tau_e$ is the charge carriers' lifetime, $D$ is the diffusion coefficient of charge carriers.

The diffusion coefficient $D$ and the lifetime of charge carriers $\tau_e$ are microscopic characteristics of a matter associated with the processes of diffusion and recombination of charge carriers.

We estimated the diffusion coefficient and charge carrier's lifetime for PbS films and PbS(I) films containing from 0.4 to 3.7 at.% I using the diffusion length $L$ for PbS equal to 1–5 μm from [28, 29]. The obtained values of the diffusion coefficient and charge carrier's lifetime (see Table 2) correlate with the values of the charge carrier's mobility (see Table 1) obtained from measurements of resistivity and Hall effect for films with different iodine contents.

As can be seen, there is a sharp decrease in the electron mobility at an iodine content in the film equal to 2.7 at.% due to the threshold increase in acceptor vacancies during iodine doping, what leads to a significant decrease (by two orders of magnitude) in the diffusion coefficient and an increase by 2-3 orders of magnitude in the carrier lifetime.

We compared the calculated values of the microscopic characteristics of the matter with the photoelectric parameters (Fig. 7). It can be seen that an increase in the iodine content in PbS films leads to nonlinear dependencies of both the microscopic characteristics and the photoelectric parameters of PbS(I) films. The sharp decrease in the diffusion coefficient and the sharp increase in the charge carriers lifetime at 2.7 at.% I are coincides with maximum on the dependences of the voltage sensitivity and specific detectability, as well as a quite large time constant ($\tau = 210$ μs).

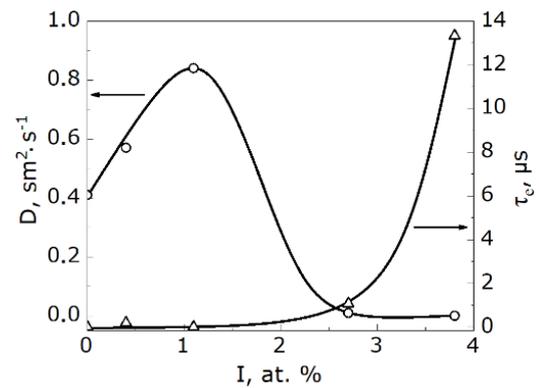

a

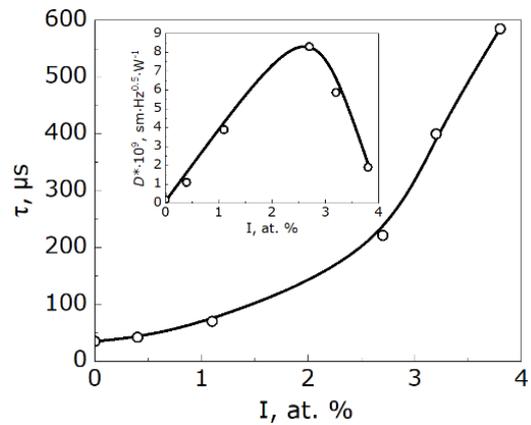

b

FIG. 7. Dependencies of time constant $\tau$ and specific detectability $D^*$ (a), diffusion coefficient $D$ and charge carriers' lifetime $\tau_e$ (b) on iodine content in PbS films.

## CONCLUSIONS

Thus, the measurements of the bulk current-voltage (I-V) characteristics and local surface I-V characteristics by the atomic force microscopy (AFM) of PbS films doped with iodine were conducted, and it was established that the bulk I-V characteristics of all studied films demonstrate a linear (ohmic) behavior. The type of local surface I-V characteristics is ohmic at the iodine content $0 < I \leq 1.1$ at.%, and becomes rectifying at $I \geq 2.7$ at.%. This is related with the features of the film surface: the height and the size of the relief elements decrease with the increase of iodine content, the surface becomes more uniform and smooth, the grain size decreases, and grain shape changes to rounded. The iodine content of 2.7 at% in PbS films is critical: the type of charge carriers changes from electrons to holes and the mobility of charge carriers decreases sharply, which is associated with the optimal concentration of acceptor vacancies that establish an equilibrium between the concentrations of electrons and holes.

An increase in the iodine content in PbS films leads to nonlinear dependencies of the microscopic characteristics and photoelectric parameters of films: there is a sharp decrease in the diffusion coefficient and the increase in the lifetime of charge carriers at 2.7 at% I in the film. Such iodine content provides maximum of voltage sensitivity and specific detectability and it is optimal for the lead sulfide film.

## ACKNOWLEDGMENTS

The work was carried out within the framework of the state assignment of the Ministry of Science and Higher Education of the Russian Federation for the IMP UB RAS.

## REFERENCES

1. Butkevich, V. G., Bochkov, V. D., & Globus, E. R. Photodetectors and photodetecting devices based on polycrystalline and epitaxial layers of lead chalcogenides. *Prikladnaya Fizika*, *6*, 66–112 (2001).
2. Kovalenko, M. v., Kaufmann, E., Pachinger, D., Roither, J., Huber, M., Stangl, J., Hesser, G., Schäffler, F., & Heiss, W. Colloidal HgTe Nanocrystals with Widely Tunable Narrow Band Gap Energies: From Telecommunications to Molecular Vibrations. Journal of the American Chemical Society, 128(11), 3516–3517 (2006).
3. Hudson Jr., R. D. Infrared System Engineering. New York, Wiley-Interscience (1969).
4. Aksenenko, M. D., Baranochnikov, M. L., & Smolin, O. V. Microelectronic photodetecting devices. Москва: Энергоатомиздат (1984).
5. Johnson, T. H. Lead Salt Detectors And Arrays PbS And PbSe. In W. L. Wolfe (Ed.), Proc. SPIE 0443, Infrared Detectors (pp. 60–94) (1983).
6. Moss, T. S. Modern Infra-red Detectors. In H. W. Thompson (Ed.), Advances in Spectroscopy: Vol. I (pp. 175–213). Interscience Publishers, Ltd (1959).
7. Bode, D. E. Lead salt detectors. In G. Hass & R.E. Thun (Eds.), Physics of Thin Films (Vol. 3, p. 275). Academic Press (1966).

8. Rogalski, A. Infrared Detectors. CRC Press (2010).
9. Gibson, A. F. The Sensitivity and Response Time of Lead Sulphide Photoconductive Cells. Proceedings of the Physical Society. Section B, 64(7), 603–615 (1951).
10. Slater, J. C. Barrier Theory of the Photoconductivity of Lead Sulfide. Physical Review, 103(6), 1631–1644 (1956).
11. Petritz, R. L. Theory of Photoconductivity in Semiconductor Films. Physical Review, 104(6), 1508–1516 (1956).
12. Neustroev, L. N., & Osipov, V. V. To the Theory of Physical Properties of Photosensitive PbS-Type Polycrystalline Films. I. Model, Conductivity and Hall Effect. Fizika i Tekhnika Poluprovodnikov, 20(1), 59–65 (1986).
13. Neustroev, L. N. Theory of the Hall Effect in a Grid of Inversion Channels. Fizika i Tekhnika Poluprovodnikov, 22(4), 773–774 (1988).
14. Dobryden, I., Touati, B., Gassoumi, A., Vomiero, A., Kamoun, N., & Almqvist, N. Morphological and electrical characterization of Cu-doped PbS thin films with AFM. Advanced Materials Letters, 8(11), 1029–1037 (2017).
15. Mikhlin, Y. L. Nonequilibrium non-stoichiometric surface layer in metal sulfide reactions. Russian Journal of General Chemistry, XLV(3), 80–85 (2001).
16. Morales-Fernández, I. E., Medina-Montes, M. I., González, L. A., Gnade, B., Quevedo-López, M. A., & Ramírez-Bon, R. Electrical behavior of p-type PbS-based metal-oxide-semiconductor thin film transistors. Thin Solid Films, 519(1), 512–516 (2010).
17. Günes, S., Fritz, K. P., Neugebauer, H., Sariciftci, N. S., Kumar, S., & Scholes, G. D. Hybrid solar cells using PbS nanoparticles. Solar Energy Materials and Solar Cells, 91(5), 420–423 (2007).
18. Maskaeva, L. N., Mostovshchikova, E. v., Markov, V. F., Voronin, V. I., Pozdin, A. v., Selyanin, I. O., & Mikhailova, A. I. Cobalt-Doped Chemically Deposited Lead-Sulfide Films. Semiconductors, 56(2), 91–100 (2022).
19. Maskaeva, L. N., Pozdin, A. v., Pavlova, A. Yu., Korkh, Yu. v., Kuznetsova, T. v., Voronin, V. I., Krivonosova, K. E., Charikova, T. B., & Markov, V. F. Charge carrier transport in PbS films doped with iodine. Physical Chemistry Chemical Physics, 26(14), 10641–10649 (2024).
20. Kaidanov V.I. and Ravich Yu.I. Deep and resonance states in $A^{IV} B^{VI}$ semiconductors //Sov. Phys. Usp., 1985, 28(1), 31–53; DOI 10.1070/PU1985v028n01ABEH003632
21. Kaidanov V.I., Nemov S.A. and Ravich Yu.I. Self-compensation of electrically active impurities by intrinsic defects in $A^{IV}B^{VI}$ semiconductors Review // Semiconductors, 1994, 28, 369–393.
22. Tsur Y., Riess I. Self-compensation in semiconductors // Phys. Rev. B 60, 8138 (1999); DOI: https://doi.org/10.1103/PhysRevB.60.8138
23. Maskaeva, L. N., Markov, V. F., Voronin, V. I., Pozdin, A. v., Borisova, E. S., & Anokhina, I. A. Structural Characteristics and Photoelectric Properties of Iodine-Doped PbS Films

Produced by Chemical Deposition. Inorganic Materials, 59(4), 349–358 (2023).
24. Rhoderick, E. H. Metal-semiconductor contacts. IEE Proceedings I Solid State and Electron Devices, 129(1), 1 (1982).
25. Lu, R. P., Kavanagh, K. L., Dixon-Warren, St. J., SpringThorpe, A. J., Streater, R., & Calder, I. Scanning spreading resistance microscopy current transport studies on doped III–V semiconductors. Journal of Vacuum Science & Technology B: Microelectronics and Nanometer Structures Processing, Measurement, and Phenomena, 20(4), 1682–1689 (2002).
26. Mishra, U. K., & Singh J. Semiconductor junctions. In Semiconductor Device Physics and Design (pp. 216–245). Springer Netherlands (2008).
27. Yafarov R.K. Effect of the built-in surface potential on the I–V characteristics of silicon mis structures. Russian Microelectronics, 48(2), 127–130 (2019).
28. Gudaev, O. A., Malinovskii, V. K., & Paul, E. E. Physical aspects of micro- and optoelectronics. Avtometriya, 4(3), 3–21 (1994).
29. Popov, V. M. Determination of the diffusion length of minority charge carriers in a semiconductor from the dynamic nonequilibrium I–V characteristics of MIS structures. Semiconductors, 48(7), 875–882 (2014).